\documentclass[preprint,showpacs,preprintnumbers,amsmath,amssymb]{revtex4}

\usepackage{graphicx}
\usepackage{dcolumn}
\usepackage{bm}

\newcommand{\be}{\begin{equation}}

\newcommand{\ee}{\end{equation}}
\newcommand{\bea}{\begin{eqnarray}}
\newcommand{\eea}{\end{eqnarray}}

\begin{document}

\preprint{BARI-TH 520/05}

\title{Effect of neutrino trapping on the three flavor FFLO phase of QCD}

\author{V.~Laporta and M.~Ruggieri}
\affiliation{Universit\`a di Bari, I-70126 Bari, Italy}
\affiliation{I.N.F.N., Sezione di Bari, I-70126 Bari, Italy}
\date{\today}

\begin{abstract}
We compute the effect of a non-zero lepton chemical potential  on
the structure of the three flavor Fulde-Ferrell-Larkin-Ovchinnikov
(FFLO) phase of QCD at finite temperature. We show that, as in the
BCS case, the lepton chemical potential favors two-species color
superconductivity and disfavors the three species pairing. We stress
that this study could be relevant for the cooling of a proto-neutron
star with a FFLO core, if the temperatures are higher than the
un-trapping temperature.
\end{abstract}


\maketitle

%%%%%%%%%%%%%%%%%%%%%%%%%%%%%%%%%%%%%%%%%%%%
%% MAINMATTER
%%%%%%%%%%%%%%%%%%%%%%%%%%%%%%%%%%%%%%%%%%%%

\section{Introduction}
Investigations on the phase diagram of Quantum-Chromo-Dynamics (QCD)
in the high quark density and low temperature regime has attracted a
lot of interest in the last years; apart the purely theoretical
speculations, these studies might be relevant for a deeper
understanding of the physics of compact stellar objects, where cold
and dense quark matter could be present. In this regime QCD predicts
Cooper pairing of quarks, due to the existence of an attractive
quark interaction in the color antisymmetric channel,
see~\cite{Collins:1974ky,Alford:1997zt} and for reviews
\cite{Rajagopal:2000wf}. Formally, one introduces a bilinear quark
expectation value (namely a di-quark condensate) in order to
describe the collective pairing; since a pair of quarks is not a
color singlet under $SU(3)_{color}$, the condensate spontaneously
breaks the color symmetry. This phenomenon is similar to the Cooper
pairing in ordinary (electromagnetic) BCS
superconductors~\cite{Bardeen:1957mv}, where the role of the quarks
is played by the electrons and $SU(3)_{color}$ is replaced by the
electromagnetic $U(1)$. Because of this analogy the quark
condensation phenomenon is known as Color Superconductivity. The
spectrum of the color superconductive phases is usually
characterized by gapped fermions, massive gluons and Goldstone
bosons related to the breaking of  some of the global symmetries of
the QCD lagrangian.

At asymptotic densities and zero temperature, the ground state of
color superconductivity with massless up ($u$), down ($d$) and
strange ($s$) quarks is the Color-Flavor-Locking (CFL)
phase~\cite{Alford:1998mk}. In the intermediate densities regime, as
could be found in the interior of a compact stellar object, one
cannot neglect neither the strange quark mass nor the differences
$\delta\mu$ in the quark chemical potentials, induced by $\beta$
equilibrium and electric and color neutrality constraints. Therefore
in the pre-asymptotic regime the CFL phase could be replaced by
another ground state. Several ground states have been proposed in
the literature as the candidates for the ``pre-CFL'' phases; we
recall here the 2SC phase~\cite{Alford:1997zt}, and the gapless
phases g2SC~\cite{Shovkovy:2003uu} and gCFL
\cite{Alford:2003fq,Alford:2004hz} (see~\cite{Ruster:2005jc} for
recent studies). In all the above phases the Cooper pairs are
characterized by a vanishing total momentum. This allows the whole
Fermi sphere to partecipate to the pairing. Moreover, the pairs have
vanishing total spin. However, the gapless phases present
chromo-magnetic
instability~\cite{Huang:2004bg,Casalbuoni:2004tb,Alford:2005qw,
Fukushima:2005cm,Fukushima:2005fh} as they show imaginary gluon
Meissner masses: this should be intimately connected to the
existence of the gapless modes in these phases. An instability is
present also in the 2SC phase~\cite{Huang:2004bg}
(in~\cite{Huang:2005pv,Hong:2005jv,Alford:2005kj,Kryjevski:2005qq,Schafer:2005ym,Gorbar:2005rx}
are discussed some antidotes to cure the chromo-magnetic instability
of the 2SC and of the gapless phases).

For appropriate values of $\delta\mu$, it can be advantageous for
quarks to form pairs with non-vanishing total momentum $2{\bf q}$.
This state, introduced for the first time in the sixties in the
contest of electromagnetic superconductors, is known as
Fulde-Ferrell-Larkin-Ovchinnikov (FFLO) phase~\cite{LOFF2}. Its
relevance in two flavor QCD has been discussed
in~\cite{Alford:2000ze,Bowers:2002xr} (see~\cite{Casalbuoni:2003wh}
for a review). By virtue of the non-vanishing pair momentum, only a
small region of the Fermi sphere is interested in the pairing
phenomenon. This results in condensates that are smaller than the
BCS ones. Moreover, the quark condensate in the FFLO phase is
space-dependent. As a consequence, the translational and rotational
symmetries are spontaneously broken and the spectrum of the low
energy excitations is enriched by the presence of the respective
Goldstone bosons, namely the phonons. As far as instability is
concerned, the authors in~\cite{Giannakis:2004pf} have shown that,
with  two flavors, the instability of 2SC implies that the FFLO
phase is energetically favored. However, the problem of the
chromo-magnetic instability of the FFLO phase is under
debate~\cite{Giannakis:2005vw,Gorbar:2005tx}.

At densities relevant for the physics of the compact stellar objects
the three flavors $u$, $d$ and $s$ can form FFLO pairs.
In~\cite{Casalbuoni:2005zp} a first attempt to the study of the
three flavor FFLO phase (at zero temperature) has been presented,
based on a Ginzburg-Landau (GL) expansion of the pressure. The
assumed pairing ansatz is
\begin{equation}
<\psi_{i\alpha}\,C\,\gamma_5\,\psi_{\beta j}> =
\sum_{I=1}^{3}\,\Delta_I({\bf r})\,\epsilon^{\alpha\beta
I}\,\epsilon_{ijI}~\label{cond}
\end{equation}with \be \Delta_I ({\bf r}) = \Delta_I
\exp\left(2i{\bf q_I}\cdot{\bf r}\right)~, \label{eq:1Ws}\ee and the
$\Delta_I$'s are the gap parameters. In the above equation $2{\bf
q_I}$ represents the momentum of the Cooper pair. For values of the
strange quark mass such that the FFLO phase is energetically favored
with respect to the homogeneous phases, and among the different
structures considered, it was found that $\Delta_1=0$ and $\Delta_2
= \Delta_3$, ${\bf q}_2 = {\bf q}_3$ is the configuration with the
highest pressure.

Recent interest has been devoted to the study of the relevance of a
lepton chemical potential on the phase diagram of
QCD~\cite{Kaplan:2001qk,Steiner:2002gx,Ruster:2005ib}. This problem
is directly connected to the phenomenon of the neutrino trapping,
which could occur in the first cooling ``era''  of a proto-neutron
star. In more detail, it is well known that the cooling of a neutron
star occurs via the emission of
neutrinos~\cite{Iwamoto:1980eb,Shapiro:1983du} (there is also a
black body contribution due to photon emission, but it is irrelevant
for our scopes); if the temperatures are not lower than $\simeq 1$
MeV then there is a spherical inner region in the star, the
neutrinosphere, from where only a very small fraction of neutrinos
escapes. The radius of the neutrinosphere (measured from the center
of the star) depends on the temperature, and decreases as the star
cools. Therefore the trapping interests regions closer and closer to
the center of the star as its temperature lowers. As a consequence
of the trapping there exists a (quasi)-conserved charge, the lepton
number, so one can introduce a lepton chemical potential $\mu_L$
associated to it. If color superconductive quark matter is present
in the cooling star, it would be  interesting to investigate about
the effect of the neutrino trapping on its structure. Recent works
along this line show that a non-zero $\mu_L$ has the effect to favor
the 2SC phase, disfavoring the CFL
phase~\cite{Steiner:2002gx,Ruster:2005ib}.

In this short note we extend the results
of~\cite{Casalbuoni:2005zp}, investigating the role of a non-zero
$\mu_L$ on the three flavor FFLO phase of QCD. Since the neutrino
trapping requires a non-zero temperature, we work at finite
temperature in all this Letter (we should mention here that the
critical temperature of the FFLO phase is expected to be lower than
the homogeneous one~\cite{Bowers:2001ip}).

\section{The model}
We consider an electrical and color neutral system of massless $u$,
$d$ and massive $s$ quarks, in $\beta$ equilibrium with massless
electrons and their neurinos. The lepton sector is described by the
Dirac lagrangian \be {\cal L}_l =
\bar{\psi}\left(i\,\partial\!\!\!\! / + \mu_l \,\gamma_0\right)\psi
~,\label{eq:LagrLept}\ee where we have collected the electron and
the neutrino fields into the doublet $\psi=(e,\nu)$, and the
chemical potential matrix is $\mu_l \equiv \text{diag}(\mu_e,
\mu_\nu) = \text{diag}(-\mu_Q + \mu_L, \mu_L)$; $\mu_Q$ is the
chemical potential associated to the conserved electric charge of
the system. From the lagrangian \eqref{eq:LagrLept} one derives the
pressure~\cite{LandauV} \be p_l = \frac{T}{2\pi^2}\sum_{a=e,\nu}
g_l\int_0^\infty {d k\, k^2 \log{(1+e^{\frac{\mu_a-k}{T}}})} ~, \ee
where $g_l$ is the degeneration factor (that counts the spin degrees
of freedom): it is equal to $2$ for electrons and $1$ for neutrinos.
One can check the correct normalization of $p_l$, that  in the limit
$T\to 0$ becomes $p_l=\mu_e^4/12\pi^2 + \mu_\nu^4/24\pi^2$.

Next we consider the quark sector. The quark lagrangian is \be {\cal
L}_q = \bar{\psi}\left(i\,\partial\!\!\!\! / -M+ \mu
\,\gamma_0\right)\psi -\frac 3 8~G~\bar\psi\gamma^\mu
\lambda_a\psi~\bar\psi\gamma^\mu \lambda_a\psi~~, \label{eq:lagr1}
\ee where $M_{ij}^{\alpha\beta} =\delta^{\alpha\beta}\, {\rm
diag}(0,0,M_s) $ is the current mass matrix;
$\mu_{\alpha\beta}^{ij}$ is the matrix of the chemical potentials:
they depend in general on $\mu$ (the baryon chemical potential),
$\mu_Q$, and $\mu_3,\,\mu_8$, related to the conserved color
charges: \be \mu_{\alpha\beta}^{ij} = \left(\mu\,\delta_{ij} +
\mu_Q\,Q_{ij}\right)\delta^{\alpha\beta} +
\left(\mu_3\,T_{3}^{\alpha\beta} +
\mu_8\,\frac{2}{\sqrt{3}}\,T_8^{\alpha\beta}\right)\delta^{ij}~,
\label{eq:potChim}\ee where $T_3$ and $T_8$ are the usual $SU(3)$
generators. Actually one should consider eight color chemical
potentials, one for each conserved color charge $n_a = <\psi T_a
\psi>$~\cite{Buballa:2005bv}; we have checked explicitly that only
$n_3$ and $n_8$ can be non-zero. This motivates the choice in
Eq.~\eqref{eq:potChim}.

The interaction term in Eq.~\eqref{eq:lagr1} is a Nambu-Jona Lasinio
inspired four fermion interaction, that mimics the one gluon
exchange of QCD. Here $G$ is a coupling constant, with dimension
{\em mass}$^{-2}$; $\lambda_a$ are color matrices and a sum over
flavors is understood. In the mean field approximation, after a
Fierz rearrangement, the interaction term becomes \be -\frac 1
2\epsilon_{\alpha\beta
I}\epsilon^{ijI}(\psi_i^\alpha\psi_j^\beta\,\Delta_I(\bm r)+\,{\rm
c.c.})\,+\,(L\to R)-\frac 1 G\Delta_I(\bm r)\Delta_I^*(\bm
r)\,,\label{8}\ee where the assumed pairing ansatz is in
Eq.~\eqref{cond}. In getting Eq.~\eqref{8} we have neglected the
chiral condensate: its effect is the dressing of the bare quark
masses; we expect that at intermediate densities (namely
$\mu\sim500$ MeV) the chiral condensate is small if compared to the
quark-quark condensate. Therefore we can safely neglect the
constituent $u$ and $d$ masses. As for the strange quark, we should
write a gap equation for its constituent mass; it should to be
solved simultaneously to the gap equations for the di-quark
condensate. A similar analysis has been performed recently
in~\cite{Ruster:2005jc}. For simplicity we do not solve the chiral
gap equation and we assume the strange quark mass as an external
parameter.

We now discuss the other approximations used in the quark sector.
First, we consider only the leading order effect of the strange
quark mass, namely a shift in its chemical potential
$\mu_s\rightarrow\mu_s - M_s^2/2\mu$. Second, to ensure color and
electrical neutrality of the system, the chemical potentials related
to the conserved charges have to satisfy the stationarity relations
\be \frac{\partial p}{\partial \mu_Q} =\frac{\partial p}{\partial
\mu_3} = \frac{\partial p}{\partial \mu_8} =0~, \label{eq:var} \ee
where $p = p_l + p_{quarks}$ (see below, Eq.~\eqref{eq:Free}). As in
Ref.~\cite{Casalbuoni:2005zp}, we work in the approximation
$\mu_3=\mu_8=0$. This assumption is justified because the phase
transition from the superconductive to the normal phase is second
order. Therefore, near the phase transition we expect $\mu_3,~\mu_8$
not too different from their values in the normal phase, namely
zero. In this way one has \be \mu_u = \mu+\frac23\mu_Q~,~~~~~ \mu_d
= \mu-\frac13\mu_Q~,~~~~~\mu_s = \mu-\frac13\mu_Q -
\frac{M_s^2}{2\mu}~. \label{eq:QuarkChem}\ee Moreover we assume
$\Delta_1 = 0$. This is justified because at $T=0$ and $\mu_L=0$ one
has $\mu_Q = - M_s^2/4\mu$. As can be inferred from
Eq.~\eqref{eq:QuarkChem} this implies that the difference of the
chemical potentials $\delta\mu_{ds}$ between $d$ and $s$ quarks is
greater than $\delta\mu_{ud}$ and $\delta\mu_{us}$. As a
consequence, the pairing between $d$ and $s$ is disfavored. This is
true also if $T\neq 0$ and $\mu_L \neq 0$ (see below). Finally, the
interaction contribution to the pressure is derived in the
Ginzburg-Landau (GL) expansion. Keeping this in mind the pressure of
quark matter is
\begin{equation}
p_q = p_0 - \left(\frac{\alpha_2}{2}\Delta_2^2 +
\frac{\beta_2}{4}\Delta_2^4 + \frac{\alpha_3}{2}\Delta_3^2 +
\frac{\beta_3}{4}\Delta_3^4 + \frac{\beta_{23}}{2}\Delta_2^2
\Delta_3^2  + O(\Delta^6)\right)~. \label{eq:Free}
\end{equation}
where $p_0$ is the pressure of the normal phase. In
Eq.~\eqref{eq:Free} the coefficients are given in terms of the
functions~\cite{Bowers:2002xr,Casalbuoni:2003wh,Casalbuoni:2005zp}
\bea
\alpha(q_,\delta\mu,T)&=&\frac{4\mu^2}{\pi^2}\left[\log\left(\frac{4\pi
T}{\Delta_0}\right) +
   {\text{Re}}\!\!\int\!\frac{d{\bf n}}{4\pi}~
\psi\left(\frac12+i\frac{\delta\mu-{\bf q}\cdot{\bf n}}{2\pi T}\right)\right] \label{eq:alphaDEF}\\
&&\nonumber\\
\beta(q_,\delta\mu,T)&=& -\frac{\mu^2}{64\pi^4
T^2}~{\text{Re}}\!\!\int\!\frac{d{\bf n}}{4\pi}~
\psi^{(2)}\left(\frac12+i\frac{\delta\mu-{\bf q}\cdot{\bf n}}{2\pi
T}\right)\label{eq:betaDEF} \eea as follows \bea  \alpha_2 =
\alpha(q_2,\frac{\mu_u-\mu_s}{2},T)~,&&~~~\alpha_3 =
\alpha(q_3,\frac{\mu_d-\mu_u}{2},T)~, \nonumber \\
\beta_2 = \beta(q_2,\frac{\mu_u-\mu_s}{2},T)~,&&~~~\beta_3 =
\beta(q_3,\frac{\mu_d-\mu_u}{2},T)~; \nonumber \eea moreover \be
\beta_{23}= -\frac{3\mu^2}{2\pi^2}\frac{1}{8\pi T^2}\int_0^1\!dy_1
dy_2\,\delta(1-y_1-y_2)\,\int_{-1}^1\!dz\,\text{Re}\,\psi^{(2)}
\left(\frac12 + \frac{i A}{2\pi T}  \right)~. \label{eq:beta23} \ee
In the above relations $\mu$ is the baryon chemical potential as
introduced in Eq.~\eqref{eq:potChim}; the functions $\psi^{(n)}(z)$
are defined as $n^{th}$-derivatives of the Euler $\psi(z)$ function,
where $\psi(z) = \Gamma'(z)/\Gamma(z)$. In Eq.~\eqref{eq:beta23} we
have introduced the function
\begin{displaymath}
A = y_1\frac{\mu_u-\mu_s}{2} - y_2\frac{\mu_d - \mu_u}{2} + z|y_1
{\bf q}_2 + y_2 {\bf q}_3|~.
\end{displaymath}
The angular integrals in Eqs.~\eqref{eq:alphaDEF},
\eqref{eq:betaDEF} and \eqref{eq:beta23} can be done exactly, but
their expressions are uninformative so we prefer to leave them in
implicit form. As for the integrals in $y_1$, $y_2$ in
Eq,~\eqref{eq:beta23}, one is performed by the $\delta(1-y_1-y_2)$
and the remaining integral can be performed numerically.

The magnitudes of the wave vectors ${\bf q}_I$ are determined by the
variational condition $\displaystyle\frac{\partial p}{\partial q_I}
= 0$; as in Ref.~\cite{Casalbuoni:2005zp}, in this paper we use this
relation at  the lowest order,
$\displaystyle\frac{\partial\alpha_I}{\partial q_I} = 0$. If $T=0$
this approximation  leads, as in the two flavor case, to the
well-known relation $q_I = 1.1997\,|\delta\mu_I|$.

\section{Results and conclusions}
\begin{figure}[t!]
\begin{center}
{\includegraphics[width=9.5cm]{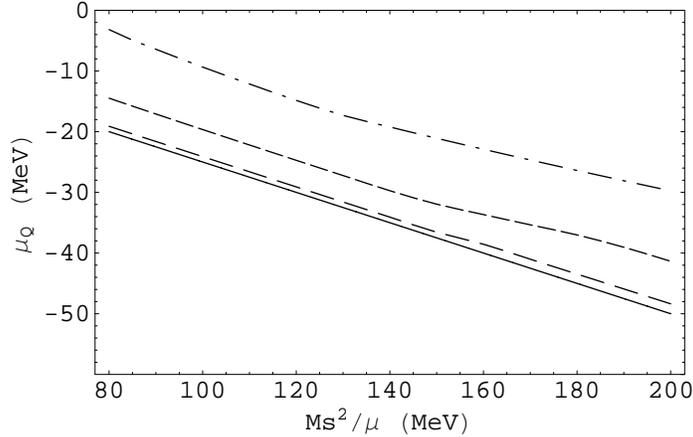}}
\end{center}
\caption{\label{FIG:mus} \footnotesize{ Chemical potential $\mu_Q$
as a function of $M_s^2/\mu$ at $T=0.1\Delta_0$. The cases $\mu_L =
0$, $100$, $200$ and $300$ MeV are represented respectively by the
solid, long-dashed, short-dashed and dot dashed lines.}}
\end{figure}

In this section we discuss our results; the coupling constant $G$
can be eliminated in favor of $\Delta_0$, the gap parameter of the
homogeneous three flavor superconductor at $T=0$ and
$\delta\mu=0$~\cite{Casalbuoni:2005zp}. We show the results obtained
for $\Delta_0 = 25$ MeV, $\mu = 500$ MeV and $T=0.1\Delta_0$ (for
different values of the parameters we get qualitatively similar
results). In Fig.~\ref{FIG:mus} we show the electrical chemical
potential $\mu_Q$ that satisfies the stationarity
condition~\eqref{eq:var} as a function of $M_s^2/\mu$ for four
values of the lepton chemical potential $\mu_L$. We notice that for
a fixed value of the strange quark mass, increasing $\mu_L$ results
in the decreasing of $|\mu_Q|$. This has important consequences on
the pairing. Indeed one has from Eq.~\eqref{eq:QuarkChem}
\begin{displaymath}
\mu_d-\mu_u = -\mu_Q~,~~~~~ \mu_u-\mu_s = \mu_Q +
\frac{M_s^2}{2\mu}~.
\end{displaymath}
For $\mu_L=0$ one has $\mu_Q=-M_s^2/4\mu$; therefore the mismatch
between the $u$ and the $d$  Fermi surfaces is the same of the $u$
and the $s$ one. For $\mu_L\neq 0$ this is no longer true: in
particular from Fig.~\ref{FIG:mus} one gets $\mu_u - \mu_s
> \mu_d - \mu_u$. This favors the pairing in the $u-d$ channel while
disfavors the $u-s$ pairing.

\begin{figure}[t!]
\begin{center}
{\includegraphics[scale=1.]{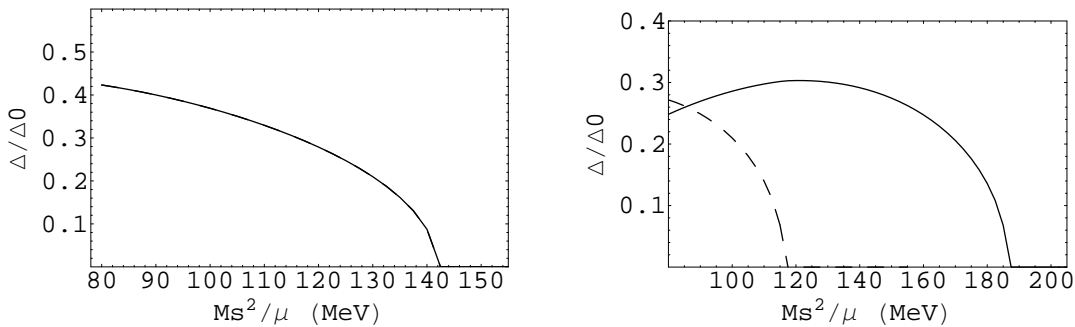}}
\end{center}
\caption{\label{FIG:gaps} \footnotesize{Left panel:
$\Delta_2/\Delta_0 = \Delta_3/\Delta_0$ against $M_s^2/\mu$,
evaluated for $\mu_L=0$. Right panel: $\Delta_2/\Delta_0$ (dashed
line) and  $\Delta_3/\Delta_0$ (solid line) as a function of
$M_s^2/\mu$, computed for $\mu_L = 200$ MeV. In both pictures
$T=0.1\Delta_0$.}}
\end{figure}

This is better expressed in Fig.~\ref{FIG:gaps}, where we show the
results for the gap parameters $\Delta_2$, $\Delta_3$ as a function
of $M_s^2/\mu$, for $\mu_L = 0$ (left panel) and $\mu_L = 200$ MeV
(right panel). In the latter case we observe that for low values of
the strange quark mass the gaps $\Delta_2$ and $\Delta_3$ have
comparable magnitude. This means that $u-d$ and $u-s$ pairing are
both active and the FFLO state is effectively a three flavor color
superconductor. Increasing $M_s$ we find the window $100 \lesssim
M_s^2/\mu \lesssim 120$ MeV where both $u-d$ and $u-s$ pairing are
active, but $\Delta_2 < \Delta_3$ meaning that the latter is
disfavored if compared to the former. At $M_s^2/\mu \simeq 120$ MeV
a second order phase transition takes place to the state with
$\Delta_2 = 0$ and $\Delta_3 \neq 0$, that is a two flavor FFLO
state with pairing in the $u-d$ channel. This is in perfect
agreement to what is found in the homogeneous
case~\cite{Steiner:2002gx,Ruster:2005ib}. Finally, for
$M_s^2/\mu\simeq185$ MeV there is a second order phase transition to
the non superconductive phase. We find also that the larger is the
value of  $\mu_L$, the larger is window of $M_s^2/\mu$ where
$\Delta_2 = 0$ and $\Delta_3\neq0$.

In conclusion, we have investigated the role of a lepton chemical
potential on the structure of the three flavor FFLO phase of QCD. We
find that a non-zero $\mu_L$ strongly favors two flavor pairing, in
the channel $u-d$, while disfavoring the pairing $u-s$. This is a
result of imposing electric neutrality in the system.

This  study could be relevant for the cooling of a proto-neutron
star. If one suppose that normal (that is, non superconductive)
quark matter is present in the star, then the neutrino emissivity
will be dominated by direct URCA
processes~\cite{Iwamoto:1980eb,Shapiro:1983du}; denoting by
$\epsilon_\nu^{URCA}(T)$ the neutrino emissivity of normal quark
matter in absence of trapping at the temperature $T$, the effect of
trapping is an exponential suppression of the emissivity
itself~\cite{Berdermann:2004da}
\begin{equation}
\bar\epsilon_\nu(r,\theta,T) =
\exp\left(-\frac{l(r,\theta)}{\lambda(T)}\right)\times\epsilon_\nu^{URCA}(T)~;
\label{eq:exp}
\end{equation}
here $\bar\epsilon_\nu(r,\theta,T)$ denotes the emissivity with the
effect of  the trapping included; $l(r,\theta)$ is the distance from
the creation point of the neutrino to the surface of the star (more
precisely, its projection along the $z$-axes).
%\begin{equation}
%l(r,\theta) = \sqrt{R^2 - r^2\sin^2\theta} - r\,\cos\theta~.
%\end{equation}
Finally, $\lambda(T)$ is the neutrino mean free path at the
temperature $T$. The exponential factor in Eq.~\eqref{eq:exp} takes
into account the probability that a neutrino created at a distance
$r$ from the center of the star can leave the star in the direction
defined by the angle $\theta$. One computes the total luminosity for
neutrino emission from the star by averaging Eq.~\eqref{eq:exp} over
all neutrino directions $\theta$ and integrating over all distances
$r$ up to the star radius.

We  turn to color superconductive quark matter. Un-trapping occurs
for temperatures  of order of $1$ MeV (for higher temperatures,
neutrinos are trapped); this imply that trapping effects should not
be neglected in emissivity computations, unless one considers the
final cooling epoch of a neutron star. Investigations on the effects
of color superconductive quark matter on the cooling evolution of a
neutron star, neglecting neutrino trapping, have been performed
in~\cite{Blaschke:1999qx,Alford:2004zr} (see also references
therein). It will be interesting to see the effect of the FFLO phase
on neutrino emissivity, if such a phase is actually present in the
core of compact stars.

\acknowledgments{We wish to thank G.~Nardulli both for having
stimulated this study and for a careful reading of the manuscript;
we thank also A.~Mirizzi for useful and enlightening discussions.
Moreover we thank the theoretical division of CERN, where this work
has been concluded, for the kind hospitality.}

\end{document}